\documentclass[aps,prb,twocolumn,superscriptaddress,groupedaddress]{revtex4}
\usepackage{graphicx}  
\usepackage{dcolumn}   
\usepackage{bm}        
\usepackage{amssymb}   
\usepackage[version=3]{mhchem} %Chemical Formulas
\usepackage{dsfont}
\usepackage{subfigure}

\DeclareMathAlphabet      {\mathbfit}{OML}{cmm}{b}{it}
\newcommand{\ket}[1]{\ensuremath{|#1\rangle}}
\newcommand{\bra}[1]{\ensuremath{\langle #1|}}

\newcommand{\diff}{\ensuremath{{\rm d}}}

\usepackage{color}

\renewcommand{\Re}{\ensuremath{{\rm Re\,}}}
\renewcommand{\Im}{\ensuremath{{\rm Im\,}}}

\begin{document}

\title{Pseudodoping of Metallic Two-Dimensional Materials}

\author{T. O. Wehling}
\affiliation{Institut f\"ur Theoretische Physik, Universit\"at Bremen, Otto-Hahn-Allee 1, 28359 Bremen, Germany}
\affiliation{Bremen Center for Computational Materials Science, Universit\"at Bremen, Am Fallturm 1a, 28359 Bremen, Germany}

\date{\today}

\begin{abstract}
We demonstrate how weak hybridization can lead to apparent heavy doping of 2d materials even in case of physisorptive binding. Combining ab-intio calculations and a generic model we show that strong reshaping of Fermi surfaces and changes in Fermi volumes on the order of several 10$\%$ can arise without actual charge transfer. This pseudodoping mechanism is very generically effective in metallic 2d materials either weakly absored to metallic substrates or embedded in vertical heterostructures. It can explain strong apparent doping of \ce{TaS2} on Au (111) observed in recent experiments. Consequences of pseudodoping for many-body instabilities are discussed.
\end{abstract}

\maketitle

\section{Introduction}
Two-dimensional electron systems host highly intriguing many-electron states including competing superconducting, nematic, magnetic, excitonic or charge ordered phases and can be controlled down to monolayer thickness \cite{novoselov_two-dimensional_2005,taniguchi_electric-field-induced_2012,ye_superconducting_2012,xi_strongly_2015,yu_gate-tunable_2015,cao_quality_2015,costanzo_gate-induced_2016}. Generally, the phase diagrams of these materials depend on carrier concentrations and doping can strongly affect their electronic properties. There is however a problem in many systems: Changes to electronic states (e.g. switching from a Mott insulator to a superconductor \cite{lee_doping_2006,yu_gate-tunable_2015}) typically require electron or hole doping on the order of a few 10$\%$ of an electron or hole per unit cell \cite{taniguchi_electric-field-induced_2012,ye_superconducting_2012,yu_gate-tunable_2015,costanzo_gate-induced_2016}. This translates into into carrier concentrations $\gtrsim 10^{14}$\,cm$^{-2}$ which are out of reach for gating in standard field effect transistor geometries but require ionic liquids or chemical means like atom substitution, intercalation of alkali atoms etc.. I.e. doping at this level is potentially related with severe chemical changes of the material or substational disorder.

\begin{figure}%
\includegraphics[width=0.9\columnwidth]{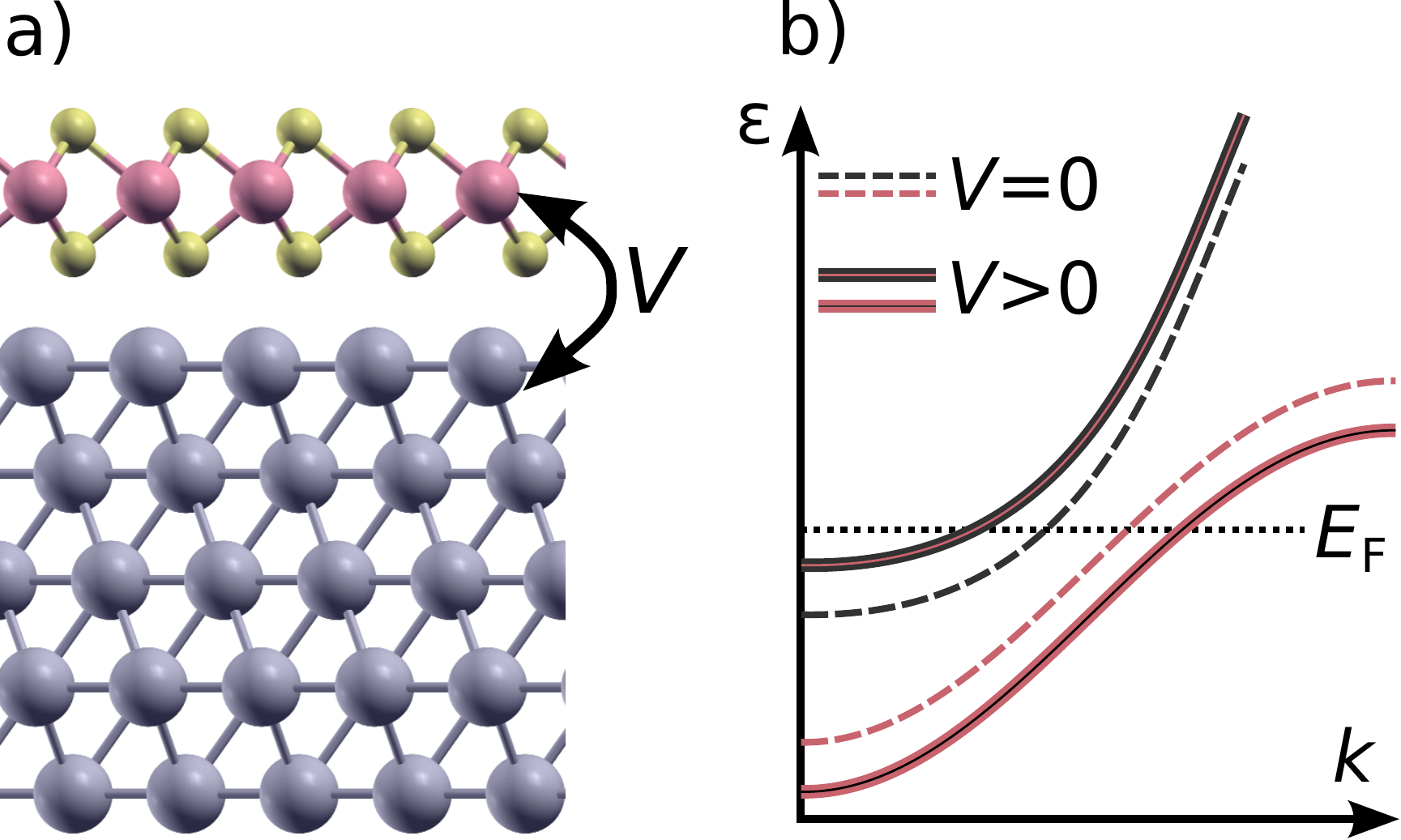}%
\caption{Coupling between a 2d material like \ce{TaS2} and a metal substrate. (a) Structure of \ce{TaS2} on Pb (111). Adsorption on the metal surface leads to hybridization $V$ between the electronic bands of the surface and the 2d material. As a consequence, there is an admixture of surface derived orbitals (gray) to the \ce{TaS2} states (purple) and level repulsion between the hybridizing bands illustrated in (b). The level repulsion leads to apparent hole-doping of the upper (surface derived) and electron doping of the lower (\ce{TaS2} derived) band.}%
\label{fig:hyb_doping_scheme}%
\end{figure}

In this paper, we discuss an alternative doping mechanism illustrated in Fig. \ref{fig:hyb_doping_scheme}. We show that hybridization of metallic 2d materials and substrates can provide effective carrier doping of more than 10$\%$ of an electron or hole per unit cell even in the case of weak physisorption. We explain how changes in Fermi volumes can arise without actual charge transfer between substrate and 2d material. We illustrate the order of magnitude of this effect with the real material example of \ce{TaS2} supported by Au and Pb metal surfaces. Afterwards the generic "pseudodoping" mechanism is explained within a simple two-band model and a self-energy formalism, which we use to disucss the impact of peudodoping on possible electronic phases. We argue that the recently observed strong doping of \ce{TaS2} on Au (111) \cite{sanders_crystalline_2016} is indeed largely hybridization induced.

\section{Realistic materials and substrates} 
We start with the example of a monolayer 1H-\ce{TaS2} absorbed to the (111) surfaces of Pb and Au. %, which are widely used in surface science experiments. 
From their interaction with graphene both Pb (111) and Au (111) are known as weakly coupling surfaces \cite{giovannetti_doping_2008,khomyakov_first-principles_2009,zhu_formation_2011}, i.e. the adsorption heights of graphene on these substrates exceed 3\AA\, which is indicative of van der Waals binding. 

\begin{figure}%
\includegraphics[width=\columnwidth]{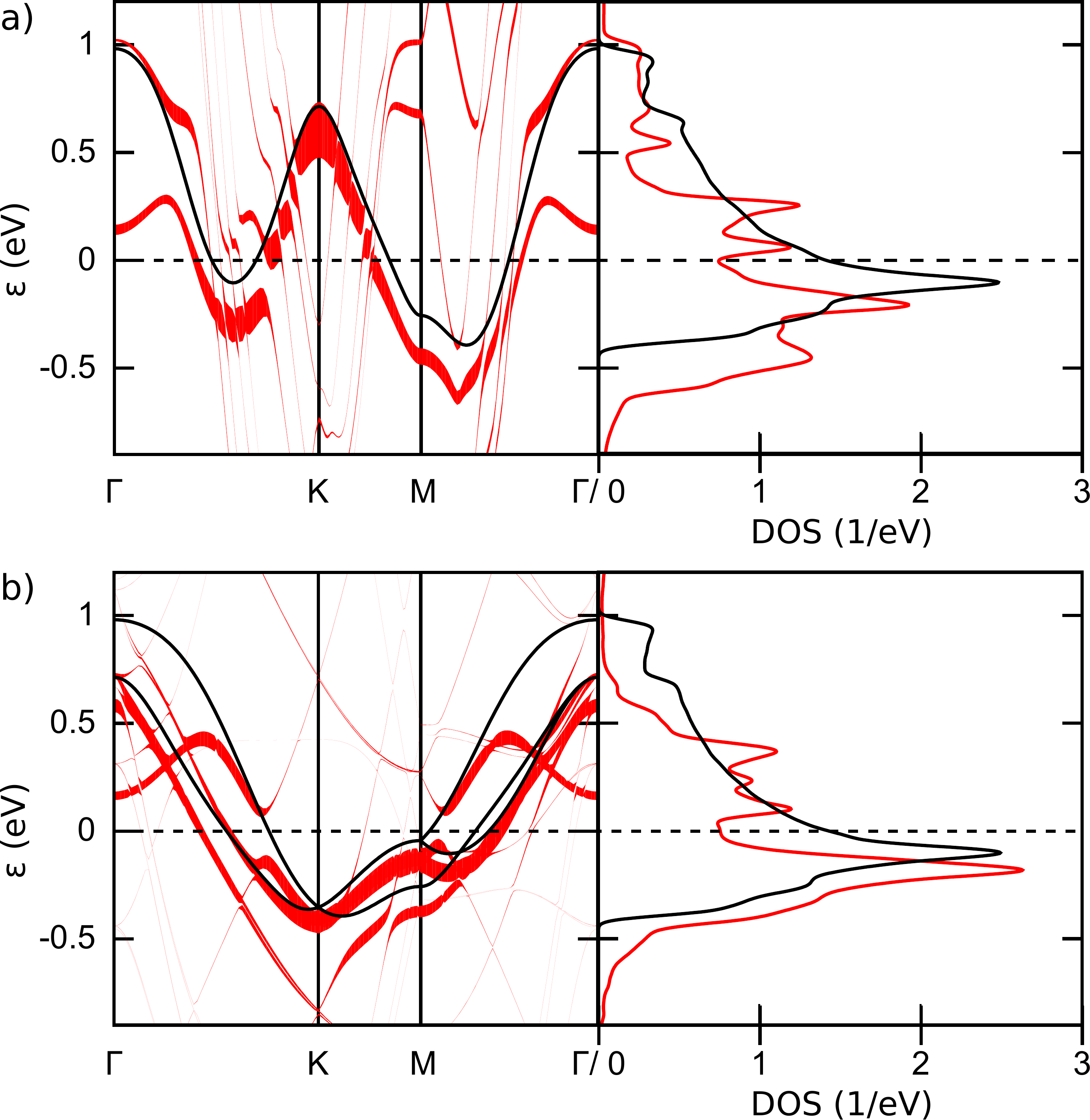}
\caption{Band structures and density of states of single layer 1H-TaS$_2$ on the Pb (111) (a) and the Au (111) surface (b). In the band structures (left panels) of the adsorbed system (red) the thickness of the lines quantifies the Ta-$d$ weight of the respective bands. The band structure of free standing monolayer 1H-\ce{TaS2} is shown in both cases in black. The corresponding density of states projected on the Ta-$d$ orbitals are shown in the right panels. All energies are given relative to the Fermi level $E_F=0.$}%
\label{fig:TaS2_surf_adsorption_doping}%
\end{figure}

To study the interaction of 1H-\ce{TaS2} with these surfaces we performed density functional calculations. (See methods section for computational details.) Under $5\%$ of compressive strain the Pb (111) surface matches the primitive unit cell of 1H-\ce{TaS2}. A comparison of the band structures of 1H-\ce{TaS2} on Pb (111) and a free standing monolayer of 1H-TaS$_2$ is shown in Fig. \ref{fig:TaS2_surf_adsorption_doping}a. The bands with sizeable Ta-$d$ weight in the adsorbed monolayer roughly follow the dispersion of free standing \ce{TaS2} in most parts of the Brillouin path. This is understandable given the adsorption height (i.e. the height of the lowest S atoms above the upmost Pb atoms) of 2.86~\AA\,, which is indicative of physisorbptive coupling between \ce{TaS2} and its substrate. 

There are however several details in which the dispersion of the adsorbed and free standing material differ from each other. In several parts of the Brillouin zone the \ce{TaS2} derived band on Pb (111) is shifted with respect to the corresponding band of free standing \ce{TaS2}. This shift amounts to $\gtrsim -0.15$~eV for most parts of the \ce{TaS2} dispersion below the Fermi level $E_F\equiv 0$ and translates via the density of states (DOS) $\rho(E_F)\approx 2$~eV$^{-1}$ into an apparent doping on the order of 0.3 electrons per \ce{TaS2} unit cell.
Considering however also the unoccupied part of the \ce{TaS2} derived bands shows that the shift relative to the free standing is strongly $k$-dependent, as is 
typical for hybridization effects. The avoided crossings in the $\Gamma$-M and $\Gamma$-K direction further substatiate that there is indeed significant hybridization between the Pb (111) surface bands and the states derived from the \ce{TaS2} layer.

A similar picture also emerges for \ce{TaS2} absorbed to Au (111). Fig. \ref{fig:TaS2_surf_adsorption_doping}b shows the band structure of $\sqrt{3}\times\sqrt{3}$\,R$30^\circ$ \ce{TaS2} supercell absorbed to a $2\times 2$ supercell of the Au (111) surface, where Au has been laterally compressed by 0.5 $\%$. Again, the \ce{TaS2} derived bands on the surface follow roughly the (backfolded) band structure of free standing \ce{TaS2} with major deviations between the two being several avoided crossings and $k$-dependent shifts in the surface band structure. The occupied part of the \ce{TaS2} derived conduction band is shifted by approximately $-0.1$\,eV relative to the corresponding band in free standing \ce{TaS2} which is in agreement with the photoemission experiments reported in \cite{sanders_crystalline_2016} and formally amounts to a doping level of 0.2 electrons per unit cell. Such a doping level is quite unexpected given that Au is rather inert. Indeed, in the case of graphene doping levels on the order of several 0.1$e^-$\, per unit cell are achieved e.g. by alkali intercalation but clearly not by adsorption on surfaces like Au (111) which merely leads to charge transfer on the order of $\lesssim 0.01e^-$\, per unit cell \cite{giovannetti_doping_2008,khomyakov_first-principles_2009}. 

Charge transfer $\Delta N$ (given in electrons per unit cell) over an effective distance $d$ is associated with an electrostatic potential difference $\Delta V=\alpha d\Delta N$, where $\alpha=e^2/\epsilon_0 A=18.6 eV/$\AA\, and $A=9.7$\AA$^2$ is the area of the \ce{TaS2} unit cell. Even if we assume that the effective distance between the charges in the 2d monolayer and the substrate is $d=1$\AA, i.e. much smaller than the typical distances of $\sim 4$\,\AA\, from the center of the \ce{TaS2} layer to the uppermost substrate atoms, a charge transfer of 0.2$e^-$ as found experimentally \cite{sanders_crystalline_2016} and theoretically in the case of 1H-\ce{TaS2} on Au (111) would translate into a potential energy difference of $\Delta V\approx 4$\,eV. Comparing the work functions of \ce{TaS2} $W_{\rm TaS_2}=5.6$~eV \cite{Shimada_JJAP94},Au (111) $W_{\rm Au}=5.31$~eV, Pb (111) $W_{\rm Pb}=4.25$~eV, \cite{Michaelson_JAP77} it is clear that potential energy differences on the order of $\Delta V\approx 4$~eV are unexpected. Indeed, estimates for the charge transfer $\Delta N$ based on work function differences \cite{giovannetti_doping_2008,khomyakov_first-principles_2009} generally arrive at $\Delta N \ll 0.1e^-$ per unit cell. We argue in the following that \textit{metallic} 2d materials are prone to "pseudodoping" and substrate induced changes in Fermi volumes, which do \textit{not} primarily relate to charge transfer but rather to hybridization.

\section{Model of pseudodoping} We consider a model involving two bands derived from orbitals $\ket{a}=(1,0)$ of the 2d material and $\ket{b}=(0,1)$ of the substrate, where we assume a (real valued) hybridization $V\geq 0$ and a constant offset $-2\Delta$ in the on-site energies between the two. The resulting Hamiltonian reads
\begin{equation}
\hat H(k)=\epsilon_0(k)\mathbf{1}+V\sigma_1-\Delta\sigma_3,
\label{eq:H_2b2}
\end{equation}
where we used the Pauli matrices $\sigma_i$ and summarized all $k$-dependence of the initial dispersion in $\epsilon_k$. The eigenstates are 
\begin{align}
\ket{-}&=\cos(\varphi)\ket{a}-\sin(\varphi)\ket{b},\nonumber\\
\ket{+}&=\cos(\varphi)\ket{b}+\sin(\varphi)\ket{a}
\end{align}
with $\tan2\varphi=V/\Delta$ and corresponding energies 
\begin{equation}
\epsilon_{\mp}(k)=\epsilon(k)\mp\sqrt{\Delta^2+V^2}.
\label{eq:eps_pm}
\end{equation}
With the Fermi energy $E_F=0$, the ground state density matrix reads thus 
\begin{equation}
\hat N(k)=\Theta(-\epsilon_-(k))\ket{-}\bra{-}+\Theta(-\epsilon_+(k))\ket{+}\bra{+}.
\end{equation}
The central point is now to contrast the occupation of the bands 
\begin{equation}
N_\mp=\int\diff^2 k \bra{\mp}\hat N(k)\ket{\mp}
\end{equation} with the occupation of the orbitals 
\begin{equation}
N_{i}=\int\diff^2 k \bra{i}\hat N(k)\ket{i},\text{where}\; i\in\{a,b\}.
\end{equation}
 The band occupations $N_\mp$ are those measured e.g. in angular resolved photoemission or quasi particle interference in scanning tunneling spectroscopy, while the orbital occupations $N_{a}$ and $N_{b}$ manifest in core level spectroscopies and also determine electric fields at the interface. Here, we have $N_a=\cos^2(\varphi)N_-+\sin^2(\varphi)N_+$ and $N_b=\cos^2(\varphi)N_++\sin^2(\varphi)N_-$.

For vanishing hybridization ($V=0 \Rightarrow \varphi=0$) the orbital occupancies simply coincide with the band occupancies: \[N^0_{a/b}=N^0_{\mp} =\int_{-\infty}^0\diff^2 \epsilon\rho^0(\epsilon\pm\Delta),\] where $\rho^0(\epsilon)=\int{\diff k}\delta(\epsilon-\epsilon_0(k))$ is the DOS associated with the dispersion $\epsilon_0(k)$. At finite $V$ the band fillings become 
\begin{align}
N_\mp&=\int_{-\infty}^0\diff \epsilon\rho^0(\epsilon\pm\sqrt{\Delta^2+V^2})\nonumber\\
&=N^0_\mp+\int_{-\infty}^0\diff \epsilon\rho^0(\epsilon\pm\sqrt{\Delta^2+V^2})-\rho^0(\epsilon\pm\Delta)\nonumber\\
%&\approx N^0_\mp\pm\int_{-\infty}^0\diff \epsilon (\partial_\epsilon \rho^0(\epsilon))(\sqrt{\Delta^2+V^2})-\Delta)\nonumber\\
&\approx N^0_\mp\pm \rho^0(0)V^2/2\Delta
\end{align}
I.e. in ARPES experiments, the upper (lower) band appear hole (electron) doped by an amount of $\rho^0(0)V^2/2\Delta$. 

However, the occupancy of the orbital $\ket{a}$ in the 2d layer is
\begin{align}
N_a&=\cos^2(\varphi)N_- + \sin^2 (\varphi)N_+\nonumber\\
%&=N_-+\sin^2 (\varphi) (N_+-N_-)\nonumber\\
%&\approx N_-+ \varphi^2 (N^0_+-N^0_-)\nonumber\\
&\approx N^0_-+\rho(0)\left[V^2/(2\Delta)+ \varphi^2 (-2\Delta)\right]\nonumber \\
%&\approx N^0_-+\rho(0)(V^2/2\Delta+ (V/2\Delta)^2 (-2\Delta)\nonumber \\
&= N^0_- + O(V^3).
\end{align}
Analogously we find $N_b\approx N^0_+$ to second order in $V$. I.e. the actual charge transfer cancels to leading order in $V$ while the apparent doping does not.

%%% Cancellation in Na/b in more general case; hybridization self-energy terms
The example discussed so far here is clearly a special one. In the general case, hybridization $V_{nk}$ of an electronic band from a 2d material with substrate bands at energies $\epsilon_{nk}$ can be captured by the self-energy (also called hybridization function)
\begin{equation}
\Sigma(k,\omega)=\sum_n\frac{|V_{n,k}|^2}{\omega+i0^+-\epsilon_{nk}}.
\label{eq:self-en}
\end{equation}
$\Re\Sigma(k,\omega)$ acts like an energy-dependent potential on the electronic states, while $\Im\Sigma(k,\omega)$ quantifies the broadening of the spectral lines. ($\omega$ denotes here the energy.) 

In the weak coupling case, we can linearize the self-energy around the Fermi level, i.e. $\Sigma(k,\omega)\approx\Sigma^0(k)+\omega\Sigma'(k)$ with
\begin{align}
\Sigma^0(k)&=-\sum_n\frac{|V_{n,k}|^2}{\epsilon_{nk}}\nonumber\\
\Sigma'(k)&=-\sum_n\frac{|V_{n,k}|^2}{(\epsilon_{nk})^2}.
\label{eq:self-en-lin}
\end{align}
$\Sigma^0(k)$ appears as an $k$-dependent energy shift, which can be naturally measured in ARPES. We have $\Sigma^0(k)<0$ ($\Sigma^0(k)>0$), where the substrate bands are predominantly above (below) the Fermi level. Correspondingly, there is an apparent electron (hole) doping of the 2d material as manifesting in a change in the band occupancy $\Delta N\approx-\rho(0)\Sigma^0(k_F)$, where $k_F$ is the Fermi momentum. Assuming that $V_{n,k}$ is on the order of interlayer couplings in 2d materials, i.e. $V_{n,k}\lesssim 300$meV, we arrive with $|\epsilon_{nk}|\sim 1$eV at the experimentally measured Fermi level shifts of $|\Sigma^0(k)|\sim 100$\,meV. 

The renormalization factor $Z_k=(1-\partial_\omega \Sigma(k,\omega))^{-1}\approx (1-\Sigma'(k))^{-1}\approx 1-\Sigma'(k)$ gives an estimate of $\ket{a}$-orbital spectral weight in the hybridized band originating from the 2d material. The remaining spectral weight $1-Z_k$ is carried by the substrate bands. The hybridization induced change in the occupation of the orbital $\ket{a}$ is thus 
\begin{equation}
\Delta N_a\approx-\rho(0)\Sigma^0(k_F)+\Sigma'(k_F)(N_s-N_a^0),
\label{eq:DeltaNa_Sigma}
\end{equation}
where $N_s$ is the occupation of the substrate states and we disregarded the $k$-dependence in $\Sigma'(k)$ for simplicity. If the substrate bands are predominantly above the band in the 2d layer, i.e. $\epsilon_a(k)<\epsilon_{nk}$, we have $\Sigma^0(k_F)<0$ and also $N_s-N_a^0<0$. Analogously, we have $\Sigma^0(k_F)>0$ and $N_s-N_a^0>0$ for $\epsilon_a(k)>\epsilon_{nk}$. Therefore, the two summands on the right hand side of Eq. (\ref{eq:DeltaNa_Sigma}) carry typically an opposite sign and tend to cancel each other at least partially. In hybrid systems consisting of a metallic 2d material on a metallic substrate, the apparent doping $\Delta N$ is thus generally stronger than the actual charge transfer $\Delta N_a$. The same line of argumentation also applies to hybrid structures made from stacking different 2d materials on top of each other.

%%% Coulomb interaction effects: Drive instabilities
\section{Pseudodoping and electronic instabilities} Interaction terms like Coulomb interactions naturally couple localized states. In the example of a Hubbard type interaction they are of the form
\begin{equation}
H_U=U\sum_{i} \hat n^a_{i\uparrow} \hat n^a_{i\downarrow},
\label{eq:H_int}
\end{equation}
where $i$ refers to lattice site of the electrons in the 2d layer, $\sigma=\{\uparrow,\downarrow\}$ to their spin and $n^a_{i\sigma}$ are the corresponding occupation number operators. $U$ is the on-site interaction matrix element and can describe repulsive ($U>0$) or attractive interaction ($U<0$). 

We assume that the substrate states are non-interacting and that their DOS at the Fermi level is small as compared to the DOS of the 2d material. Then, the admixture of the substrate states to the bands of the 2d material upon hybridization reduces the \textit{effective} interaction inside the hybridized band by a factor of $Z^2$, i.e. $U\to U^{\rm eff}=Z^2 U$. For weak coupling instabilities such as BCS superconductivity, characteristic transition temperatures $T_C\sim\exp[-1/U^{\rm eff}\rho]$ are determined by the interaction $U^{\rm eff}$ and DOS at the Fermi level $\rho$ resulting from the hybridized band. If the DOS of the original 2d band is structureless, we expect simply a reduction of $T_C$. An analogous line of argumentation holds for weak coupling charge- or spin-density wave instabilities at some wave vector $Q$, where the susceptibility $\chi(Q)=\int\diff^2 k (f(\epsilon_k)-f(\epsilon_{k+Q}))/(\epsilon_k-\epsilon_{k+Q})$ plays the role of an effective density of states. Thus, hybridization effects should quench tendencies towards weak coupling electronic instabilities, if the DOS/ generalized susceptibilities are essentially independent of the Fermi energy. An even stronger suppression of instabilities is expected if $\rho$ or $\chi(Q)$ are reduced upon hybridization related pseudodoping. 
A comparison of the DOS of 1H TaS$_2$ on Au (111) and in its free standing form (c.f. Fig. \ref{fig:TaS2_surf_adsorption_doping}) shows that there is indeed a reduction of $\rho(E_F)$ upon deposition on Au. I.e. both, the reduction of effective coupling constants and the reduced DOS, will contribute here to the suppression of charge density wave / superconducting states observed in Ref. \onlinecite{sanders_crystalline_2016}.

If, however, hybridization shifts a strong peak in the DOS of the 2d material's band towards $E_F$ such that the hybridization induced increase in $\rho$ overcompensates the reduction of the effective interaction by the factor $Z^2$, we arrive at an increase in $T_C$. 

\section{Conclusions} We have shown that hybrization of metallic 2d materials with their substrates can lead to apparent doping manifesting as shifts of bands in ARPES experiments. This "pseuddoping" can involve much less actual charge transfer between the layer and its substrate than changes in the Fermi surface would suggest. Nonetheless, instabilities of the electronic system towards symmetry broken states, particularly weak coupling instabilities, are expected to be highly sensitive to this kind of doping. The pseudodoping mechanism outlined, here, might explain recent ARPES experiments reporting strong doping of \ce{TaS2} on Au (111) \cite{sanders_crystalline_2016}.

%%%BCS M~Z + changes in rho, where rho follows Delta N, i.e. pseudodoping affects rho as entering BCS theory. 
%%%Slater Antiferromagnetism: Same phenomenology holds. Diverging chi(Q).
%%%Strong coupling: Situation more complicated: Local moment formation essentially unaffected hybridization doping. Mott insulator cannot form. metallic substrate states will penerate into 2d layer via Kondo effect. 

\section{Acknowledgments}
This work was supported by the European Graphene Flagship. The numerical computations were carried out on the Norddeutscher Verbund zur F\"orderung des Hoch- und H\"ochstleistungsrechnens (HLRN) cluster. We thank P. Hofmann and C. Sanders for discussions of the ARPES experiments of Ref. \cite{sanders_crystalline_2016}.

\section{Methods}
We performed density function theory calculations using the Vienna Ab Initio Simulation Package (VASP) \cite{Kresse:PP_VASP} with the projector augmented wave basis sets \cite{Bloechl:PAW1994,Kresse:PAW_VASP} and the generalized gradient approximation to the exchange correlation potential \cite{PBE_96}. In all cases we fixed the in-plane lattice constant of 1H-\ce{TaS2} to the experimental value of $a=3.316$\AA\; \cite{mattheiss_band_1973}. 

The Pb (111) and the Au (111) surfaces where modeled using slabs with at thickness of 5 atomic layers and a single layer of 1H-\ce{TaS2} absorbed on the upper side of the slabs (see Fig. \ref{fig:hyb_doping_scheme}a). The Au (111) slab was furthermore terminated with H on the bottom side of the slab, i.e. the side without \ce{TaS2} coverage. The lateral coordinates of all atoms were kept fixed and we laterally compressed the Pb (111) surface (Au (111) surface) by $5\%$ ($0.5\%$) to match a $1\times 1$ (($\sqrt{3}\times\sqrt{3}$)R$30^\circ$) unit cell of 1H-\ce{TaS2} with a $1 \times 1$ ($2\times 2$) unit cell of the metal surface. The vertical coordinates of the Ta and S atoms were relaxed until forces acting on them were below {0.01~eV/\AA} leading to structures, where the closest vertical distance between S-atoms and Pb (Au) surface atoms is 2.86~\AA\, (2.86~\AA).

\bibliography{Hyb_doping}
\bibliographystyle{apsrev}
\end{document}